\documentclass[reqno,11pt]{amsart}

\usepackage{amsmath}
\usepackage{amssymb}

\newtheorem{Theorem}{Theorem}[section]
\newtheorem{Lemma}[Theorem]{Lemma}

\newcommand{\thref}[1]{Theorem \ref{#1}}
\newcommand{\leref}[1]{Lemma \ref{#1}}

\theoremstyle{definition}
\newtheorem{Remark}[Theorem]{Remark}

\newtheorem*{Remark*}{Remark}

\numberwithin{equation}{section}

\begin{document}

\newcommand{\pd}{\partial}
\newcommand{\res}{\mathrm{res}}
\newcommand{\alg}{\mathrm{alg}}
\newcommand{\Gr}{\mathrm{Gr}}
\newcommand{\C}{\mathbb{C}}
\newcommand{\cA}{\mathfrak{A}}
\newcommand{\cL}{\mathcal{L}}
\newcommand{\cM}{\mathcal{M}}

\title{Finite heat kernel expansions on the real line}

\author[P.~Iliev]{Plamen~Iliev}
\address{School of Mathematics, Georgia Institute of Technology, 
Atlanta, GA 30332--0160, USA}
\email{iliev@math.gatech.edu}

\date{May 12, 2005}


\begin{abstract} Let $\cL=d^2/dx^2+u(x)$ be the one-dimensional Schr\"odinger 
operator and $H(x,y,t)$ be the corresponding heat kernel. We prove that the 
$n$th Hadamard's coefficient $H_n(x,y)$ is equal to $0$ if and only if 
there exists a differential  operator $\cM$ of order $2n-1$ such that 
$\cL^{2n-1}=\cM^2$. Thus, 
the heat expansion is finite if and only if the potential $u(x)$ is a 
rational solution of the KdV hierarchy decaying at infinity studied in 
\cite{AM,AMM}. Equivalently, one can characterize the corresponding operators 
$\cL$ as the rank one bispectral family in \cite{DG}.
\end{abstract}

\maketitle

\section{Introduction and description of the main result} 

Let $u(x)$ be a smooth function and let $\cL$ be the one-dimensional 
Schr\"odinger operator
\begin{equation}
\cL=\frac{d^2}{d x^2}+u(x).
\end{equation}
The heat kernel $H(x,y,t)$ is the fundamental solution of the heat equation
\begin{equation}\label{1.2}
\left(\frac{\pd}{\pd t}-\cL\right)f=0.
\end{equation}
It is well known that $H(x,y,t)$ has an asymptotic expansion of the form
\begin{equation}\label{1.3}
H(x,y,t) \sim \frac {e^{-\frac{(x-y)^2}{4t}}}{\sqrt{4\pi t}} \left( 1 +
\sum_{n=1}^{\infty} H_n(x,y)t^n\right) \text{ as }t\rightarrow 0+,
\end{equation}
valid for $x$ close to $y$. The coefficients $H_n$ are the so called 
Hadamard's coefficients 
\cite{Hadamard}. 

In a series of papers 
Berest and Veselov studied linear hyperbolic partial differential operators 
that satisfy the Huygens principle and discovered a beautiful connection 
with integrable systems and the bispectral problem \cite{DG}, see for example 
\cite{Be,BV} and the references therein. Gr\"unbaum \cite{Gr} observed that
a similar connection is present in the context of the heat equation. More 
precisely, he showed that for the first few potentials 
$u_i(x)=2\log(\tau_i(x))''$, with $\tau_0=1$, $\tau_1(x)=x$, 
$\tau_2(x)=x^3/3-s_3$, formed by the rational solutions of Korteweg-de Vries 
(KdV) equation, the asymptotic expansion \eqref{1.3} gives rise to an exact 
formula consisting of finite number of terms. The corresponding operators 
$\cL$ belong to the rank one family of solutions to the bispectral problem 
\cite{DG}.

The main result of the present paper is to prove that this holds for all
rank one bispectral operators and that this completely characterizes 
the rank one family. More precisely we have 

\begin{Theorem} \label{th1.1}
The following conditions are equivalent
\begin{itemize}
\item[(i)] $H_n(x,y)=0$ for all $x$ and $y$;
\item[(ii)] There exists a differential operator $\cM$ of order $2n-1$ such 
that
\begin{equation}\label{1.4}
\cL^{2n-1}=\cM^2.
\end{equation}
\end{itemize}
\end{Theorem}
Operators $\cL$ and $\cM$, satisfying  \eqref{1.4} were studied in 
a paper by Burchnall and Chaundy\footnote{They 
considered even a more general situation of operators $P$ and $Q$ of arbitrary 
orders, satisfying $P^{\deg(Q)}=Q^{\deg(P)}$.} \cite{BC}. 
Their main result essentially means that the 
operator $\cL$ can be obtained by applying a sequence of rational Darboux 
transformations to the differential operator $\cL_0=d^2/dx^2$. According to 
a result by Adler and Moser \cite{AM} the corresponding potentials $u(x)$ are 
precisely the rational solutions of the KdV equation decaying at $\infty$. 
The latter were discovered by Airault, McKean and Moser \cite{AMM} and 
mysteriously connected with the Calogero-Moser system.  
Finally, the operators $\cL$ in (ii) coincide with the rank one 
bispectral family in the work of Duistermaat and Gr\"unbaum \cite{DG}. 
Thus, the present paper adds one more characterization of the operators 
$\cL$ satisfying condition (ii), namely the finiteness property of the heat 
kernel expansion.

The proof of this theorem can be obtained using the connection between the 
heat kernel and the KdV hierarchy. This goes back to the 
pioneer work of McKean and van Moerbeke \cite{McKvM}, where it was 
shown that, restricted on the diagonal $x=y$, the Hadamard's coefficients give 
the flows of the KdV hierarchy. See also \cite{ASch} 
and the references therein for extensions to matrix-valued heat kernel 
expansions and applications. 

In a recent paper \cite{I} we showed that one can go also in the opposite 
direction and construct the heat kernel using the $\tau$-function (in the 
sense of Sato see \cite{SS,Hirota}). We combine this formula for $H_n(x,y)$ 
in terms of the Baker function, the explicit description of the coordinate 
ring for the algebraic curves and the bilinear (Hirota) equations to show 
that the vanishing of the Hadamard's coefficient $H_n(x,y)$ is equivalent 
to condition (ii) in the theorem.

For related results concerning the finiteness property of the heat kernel 
expansion on the integers and rational solutions of the Toda lattice hierarchy 
see \cite{GI}. For solitons of the Toda lattice and purely discrete versions 
of the heat kernel see \cite{Haine}.

\section{Baker functions and an infinite Grassmannian}
In this section we collect some preliminary facts about the Segal-Wilson 
Grassmannian \cite{SW} and the parametrization of rank one commutative 
rings of differential operators. We also present the formula from \cite{I} 
which relates $H_n(x,y)$ with Sato theory for the KdV hierarchy. 
Finally we prove a simple lemma which allows us later to use the 
algebro-geometrical data.

Let us denote by $S^1$ the unit circle $S^1=\{z\in\C:|z|=1\}$ and let 
$H$ denote the Hilbert space $L^2(S^1,\C)$. We split $H$ as the orthogonal 
direct sum $H=H_+\oplus H_-$, where $H_+$ and (resp. $H_-$) consists of the 
functions whose Fourier series involves only nonnegative (resp. only negative) 
powers of $z$. We denote by $\Gr$ the Segal-Wilson 
Grassmannian of all closed subspaces $W$ of $H$ such that
\begin{itemize}
\item the projection $W\rightarrow H_+$ is a Fredholm operator of index zero;
\item the projection $W\rightarrow H_-$ is a compact operator.
\end{itemize}
Let $\Gr^{(2)}$ be the subspace of $\Gr$ given by 
$$\Gr^{(2)}=\{W\in\Gr:z^2W\subset W\}.$$
To each space $W\in\Gr$ there is a unique (Baker) function $\Psi_W(x,z)$ 
characterized by the following properties:
\begin{itemize}
\item $\Psi_W$ has the form 
$\Psi_W(x,z)=\left(1+\sum_{i=1}^{\infty}\psi_i(x)z^{-i}\right)e^{xz}$;
\item $\Psi_W(x,.)$ belongs to $W$ for each $x$.
\end{itemize}

We  denote by $W^{\alg}$ the subspace of  elements of finite order, i.e. 
the elements of the form $\sum_{j\leq k}a_jz^j$. This is a dense subspace of 
$W$. A very important object connected to the plane $W$ is the ring $A_W$ of 
analytic functions $f(z)$ on $S^1$ that leave $W^{\alg}$ invariant, i.e.
$$A_W=\{ f(z)\text{ analytic on }S^1:f(z)W^{\alg}\subset W^{\alg}\}.$$
For each $f(z)\in A_W$ one can show that there exists a unique 
differential operator $L_f(x,\pd/\pd x)$ such that
$$L_f\Psi(x,z)=f(z)\Psi_W(x,z).$$
The order of the operator $L_f$ is equal to the order of $f$. In general, 
$A_W$ is trivial, i.e. $A_W=\C$. The spaces $W$ which arise from 
algebro-geometrical data (which are important for us) are precisely those such 
that $A_W$ contains an element of each sufficiently large order. The 
commutative ring of differential operators
\begin{equation*}
\cA_W=\{L_f:f\in A_W\}
\end{equation*}
is isomorphic to $A_W$. Every rank one commutative ring\footnote{A commutative 
ring of differential operators is called rank one if it contains an operator 
of every sufficiently large order.}
of differential operators can be obtained by this construction for an 
appropriate $W\in\Gr$.
Notice that if $W\in\Gr^{(2)}$ then $z^2\in W$ and $\cL=L_{z^2}=d^2/dx^2+u(x)$ 
is a second-order operator.

\begin{Remark} In Sato's theory of the Kadomtsev-Petviashvili hierarchy one 
usually writes the Baker (wave) function 
$$\Psi_W(s,z)=\left(1+\sum_{i=1}^{\infty}\psi_i(s)z^{-i}\right)
e^{\sum_{k=1}^{\infty}s_kz^k}$$
depending on all times $s_1=x,s_2,s_3,\dots$. Then if $W\in\Gr^{(2)}$ 
the operator $\cL=L_{z^2}$ satisfies the KdV hierarchy
\begin{equation}\label{2.1}
\frac{\pd\cL}{\pd s_k}=[(\cL^{k/2})_+\, ,\cL], \qquad k=1,3,5,\dots,
\end{equation}
where $(\cL^{k/2})_+$ is the differential part of the pseudo-differential 
operator $\cL^{k/2}$.
\end{Remark}

On the Hilbert space $H$ we have the non-degenerate continuous skew-symmetric 
bilinear form
\begin{equation}\label{2.2}
\langle f,g\rangle=\frac{1}{2\pi i}\oint_{S^1}f(z)g(-z)dz.
\end{equation}
If $W\in\Gr$ then its annihilator
$$W^*=\{f\in H:\langle f,g\rangle=0\text{ for all }g\in W\}$$
also belongs to $\Gr$. 
The adjoint Baker function $\Psi^*(x,z)$ is defined by
$$\Psi_W^*(x,z)=\Psi_{W^*}(x,-z).$$
It follows immediately from the definition that the following bilinear 
identity 
\begin{equation}\label{2.3}
\oint_{S^1}\Psi_W(x,z)\Psi_W^*(y,z)dz =0
\end{equation}
holds (see \cite{DJKM}). 

The main result of \cite{I} is the following formula for $H_n(x,y)$
\begin{equation}\label{2.5}
H_n(x,y)=-\frac{1}{\pi i (x-y)^{2n-1}}
\oint_{S^1}g_n\left(2(x-y)z\right)\Psi_W(x,z)\Psi^*_W(y,z)dz,
\end{equation}
where 
\begin{equation}
g_n(z)=(-1)^{n-1}\left[\sum_{k=0}^{n-1}\frac{(2n-k-2)!}{k!(n-k-1)!}z^k\right]
   e^{-\frac{z}{2}}.
\end{equation}
It is interesting to notice that the function $g_n(z)$ is closely related 
to a specific Laguerre polynomial. Recall that the Laguerre polynomials 
$L_{n}^{\alpha}(z)$ are defined by the following formula
\begin{equation}
L_n^{\alpha}(z)=\sum_{k=0}^n\frac{\Gamma(n+\alpha+1)}{\Gamma(k+\alpha+1)}
\frac{(-z)^{k}}{k!(n-k)!},
\end{equation}
where $\Gamma(z)$ is the usual gamma function, see for example 
\cite[p. 77]{Leb}. 
Thus we can write $g_n(z)$ as 
\begin{equation}\label{2.8}
g_n(z)=(n-1)! L^{-2n+1}_{n-1}(z) e^{-\frac{z}{2}}.
\end{equation}
Using the above formula we prove the following lemma.

\begin{Lemma}\label{le2.2}
Let
\begin{equation}\label{2.9}
g_n(z)=\sum_{k=0}^{\infty}\beta_kz^k,
\end{equation}
be the Taylor expansion of the function $g_n(z)$ around $z=0$. Then
$\beta_{2j-1}=0$ for $1\leq j<n$
\end{Lemma}

\begin{proof}
It is well known that the function $f(z)=L_n^{\alpha}(z)e^{-\frac{z}{2}}$
satisfies the differential equation
$$zf''+(\alpha+1)f'+\left(n+\frac{\alpha+1}{2}\right)f=\frac{zf}{4},$$
see for example \cite[p. 85]{Leb}.

Using the above equation and \eqref{2.8} we see that $g_n(z)$ must satisfy 
the differential equation
\begin{equation*}
zg''_n-2(n-1)g'_n=\frac{zg_n}{4},
\end{equation*}
from which it follows immediately that 
\begin{equation}\label{2.10}
8(2j-1)(j-n)\beta_{2j-1}=\beta_{2j-3}.
\end{equation}
Combining this with $\beta_1=0$ we obtain the assertion of the lemma. 
\end{proof}

\section{Proof of the theorem}
The implication (i)$\Rightarrow$(ii) can be proved using the connection 
between Hadamard's coefficients on the diagonal and the KdV hierarchy 
\eqref{2.1}. Indeed we have 
\begin{equation}\label{3.1}
\frac{\pd H_k(x,x)}{\pd x}=\frac{2^{k-1}}{(2k-1)!!}\,
[(\cL^{\frac{2k-1}{2}})_+,\cL],
\end{equation}
see for example \cite[Theorem 5.2]{Sch}. 
If $H_n(x,y)=0$, then we have $H_k(x,y)=0$ for all 
$k\geq n$. Formula \eqref{3.1} combined with \eqref{2.1} shows that $u$ is 
stationary under the flows $\pd /\pd s_{2k-1}$ for $k\geq n$. This 
implies that $A_W\supset\C[z^2,z^{2n-1}].$
Thus the operator $\cM=L_{z^{2n-1}}$ satisfies $\cM^2=\cL^{2n-1}$, which 
gives (ii).

Below we prove the implication (ii) $\Rightarrow$ (i). If (ii) holds 
then the operator $\cL$ commutes with the operator $\cM$ of order $2n-1$, i.e. 
it belongs to a rank one commutative ring of differential operators $\cA_W$
for some plane $W\in\Gr^{(2)}$. Moreover we have
\begin{align}
\cL(x,\pd/\pd x)\Psi_W(x,z)&=z^2\Psi_W(x,z)\\
\cM(x,\pd/\pd x)\Psi_W(x,z)&=z^{2n-1}\Psi_W(x,z),
\end{align}
or equivalently
\begin{equation}
\C[z^2,z^{2n-1}]\subset A_W.
\end{equation}
This combined with \eqref{2.3} means that 
\begin{equation*}
\oint_{S^1}z^k\Psi_W(x,z)\Psi_W^*(y,z)dz=0
\text{ if }
\left\{
\text{\begin{tabular}{l}
$k$ is even or\\
$k\geq 2n-1$
\end{tabular}} 
\right.
\end{equation*}
On the other hand, \leref{le2.2} says that the first odd power of $z$ in 
the Taylor expansion of $g_n(z)$ is $z^{2n-1}$. Thus formula \eqref{2.5} 
shows that $H_n(x,y)=0$.

\begin{Remark} 
The subspaces $W$ parametrizing the operators $\cL$ described in 
\thref{th1.1}(ii) form a sub-Grassmannian of $\Gr^{(2)}$ denoted by 
$\Gr^{(2)}_0$ in \cite{SW}. It also parametrizes all rank one commutative 
rings of differential operators $\cA$ with unicursal spectral curve Spec$\cA$, 
containing an operator of order two. It is well known (see \cite[p.~46]{SW}) 
that it has a cell decomposition with cells indexed by the sets
$$S_n=\{-n,-n+2,-n+4,\dots,n,n+1,n+2,\dots\}.$$
The corresponding cell $C_n$ in $\Gr^{(2)}_0$ consists of the solutions 
to the KdV hierarchy \eqref{2.1} flowing out of the initial value
$$u(x,0,0,\dots)=-\frac{n(n+1)}{x^2}.$$
These solutions exhaust all rational solutions to the KdV hierarchy 
that vanish at $x=\infty$ \cite{AM,AMM}.
The statement of \thref{th1.1} can be reformulated as follows:
$H_{n+1}(x,y)=0$ if and only if $W\in C_k$ for some $k\leq n$.
\end{Remark}

{\bf Acknowledgments.} I would like to thank Alberto Gr\"unbaum for his 
collaboration in \cite{GI} which inspired my interest in the subject.
I also want to thank George Wilson for suggesting this simple proof of the 
implication (i)$\Rightarrow$(ii).


\end{document}